# Steganography
# An Art of Hiding Data

Shashikala Channalli, Ajay Jadhav
Sinhgad College of Engineering, Pune.

*Abstract:* In today's world the art of sending & displaying the hidden information especially in public places, has received more attention and faced many challenges. Therefore, different methods have been proposed so far for hiding information in different cover media. In this paper a method for hiding of information on the billboard display is presented. It is well known that encryption provides secure channels for communicating entities. However, due to lack of covertness on these channels, an eavesdropper can identify encrypted streams through statistical tests and capture them for further cryptanalysis. In this paper we propose a new form of steganography, on-line hiding of information on the output screens of the instrument. This method can be used for announcing a secret message in public place. It can be extended to other means such as electronic advertising board around sports stadium, railway station or airport. This method of steganography is very similar to image steganography and video steganography. Private marking system using symmetric key steganography technique and LSB technique is used here for hiding the secret information.

## I. INTRODUCTION

With the development of computer and expanding its use in different areas of life and work, the issue of information security has become increasingly important. One of the grounds discussed in information security is the exchange of information through the cover media. To this end, different methods such as cryptography, steganography, coding, etc have been used. The method of steganography is among the methods that have received attention in recent years.[1] The main goal of steganography is to hide information in the other cover media so that other person will not notice the presence of the information. This is a major distinction between this method and the other methods of covert exchange of information because, for example, in cryptography, the individuals notice the information by seeing the coded information but they will not be able to comprehend the information. However, in steganography, the existence of the information in the sources will not be noticed at all. Most steganography jobs have been carried out on images, video clips ,texts, music and sounds .Nowadays, using a combination of steganography and the other methods, information security has improved considerably. In addition to being used in the covert exchange of information, steganography is used in other grounds such as copyright, preventing e-document forging.

Steganography is derived from the Greek for covered writing and essentially means "to hide in plain sight". Steganography is the art of inconspicuously hiding data within data. The main goal of steganography is to hide information well enough such that the unintended recipients do not suspect the steganographic medium of containing hidden data Simple steganographic techniques have been in use for hundreds of years, but with the increasing use of files in an electronic format new techniques for information hiding have become possible. Most steganography jobs have been carried out on different storage cover media like text, image, audio or video. Steganography [2] & encryption are both used to ensure data confidentiality However the main difference between them is that with encryption anybody can see that both parties are communicating in secret. Steganography hides the existence of a secret message and in the best case nobody can see that both parties are communicating in secret. This makes steganography suitable for some tasks for which encryption aren't, such as copyright marking. Table 1 shows a comparison of different techniques for communicating in secret [4]. Encryption allows secure communication requiring a key to read the information. An attacker cannot remove the encryption but it is relatively easy to modify the file, making it unreadable for the intended recipient.

TABLE 1
COMPARISON OF SECRET COMMUNICATION TECHNIQUES.

| Secret Communication Techniques | Confidentiality | Integrity | Un removability |
|---|---|---|---|
| Encryption | Yes | No | Yes |
| Digital Signatures | No | Yes | No |
| Steganography | Yes/No | Yes/No | Yes |

## II. REQUIREMENTS OF HIDING INFORMATION DIGITALLY

There are many different protocols and embedding techniques that enable us to hide data in a given object. However, all of the protocols and techniques must satisfy a number of requirements so that steganography can be applied correctly [4].

The following is a list of main requirements that steganography techniques must satisfy:
a) The integrity of the hidden information after it has been embedded inside the stego object must be correct..
b) The stego object must remain unchanged or almost unchanged to the naked eye.
c) In watermarking, changes in the stego object must have no effect on the watermark.
d) Finally, we always assume that the attacker knows that there is hidden information inside the stego object.

*A. Embedding and detecting secret information*

Figure 1 shows a simple representation of the generic embedding and decoding process in steganography. In this example, a secret image is being embedded inside a cover image to produce the stego image [5]. The first step in embedding and hiding information is to pass both the secret message and the cover message into the encoder. Inside the encoder, one or several protocols will be implemented to embed the secret information into the cover message.

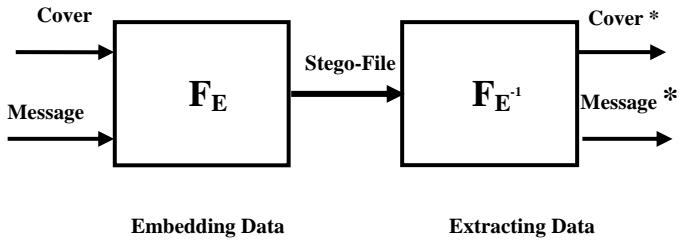

Figure . 1  Structure of Steganography System

Having produced the stego object, it will then be sent off via some communications channel, such as email, to the intended recipient for decoding. The recipient must decode the stego object in order for them to view the secret information. The decoding process is simply the reverse of the encoding process. It is the extraction of secret data from a stego object. In the decoding process, the stego object is fed in to the system. The public or private key that can decode the original key that is used inside the encoding process is also needed so that the secret information can be decoded.

After the decoding process is completed, the secret information embedded in the stego object can then be extracted and viewed.

### III. TYPES OF STEGANOGRAPHY

Steganography can be split into two types :
a) Fragile:This steganography involves embedding information into a file which is destroyed if the file is modified.
b) Robust: Robust marking aims to embed information into a file which cannot easily be destroyed.

The comparison of various methods of steganography and their advantages is as shown in the Table 2.

TABLE 2
COMPARISON OF VARIOUS METHODS OF STEGANOGRAPHY

| Sr. No | Steganography Techniques | Cover Media | Embedding Technique | Advantages |
|---|---|---|---|---|
| 1. | Binary File Technique | Binary File | watermark can be embedded by making changes to the binary code that does not affect the execution of the file | Simple to implement |
| 2. | Text Technique | Document | To embed information inside a document we can simply alter some of its characteristics.i.e. either the text formatting or characteristics of the characters | Alterations not visible to the human eye |
| 3. | Image Hiding: 1) LSB ( Least Significant Bit | Image | It works by using the least significant bits of each pixel in one image to hide the most significant bits of another. | Simple & easiest way of hiding information. |
|  | 2) DCT ( Direct Cosine Transform ) |  | Embeds the information by altering the transformed DCT coefficients. | Hidden data can be distributed more evenly over the whole image in such a way as to make it more robust. |
|  | 3) Wavelet Transform |  | This technique works by taking many wavelets to encode a whole image | Coefficients of the wavelets are altered with the noise within tolerable levels |
| 4 | Sound Technique | MP3 files | Encode data as a binary sequence which sounds like noise but which can be recognised by a receiver with the correct key | Used for watermarking by matching the narrow bandwidth of the embedded data to the large bandwidth of the medium |
| 5. | Video Technique | Video Files | A combination of sound and image techniques can be used | The scope for adding lots of data is much greater |

## IV PROPOSED WORK

The main goal of this method is to hide information on the output image of the instrument (such as image displayed by an electronic advertising billboard). This method can be used for announcing a secret message in a public place. In general, this method is a kind of steganography, but it is done in real time on the output of a device such as electronic billboard. Following are the steps involved in embedding the secret information within a cover media.

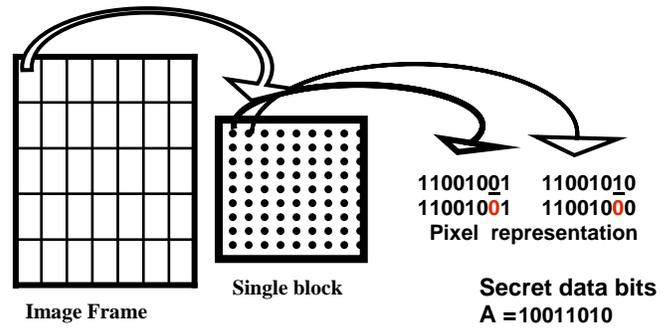

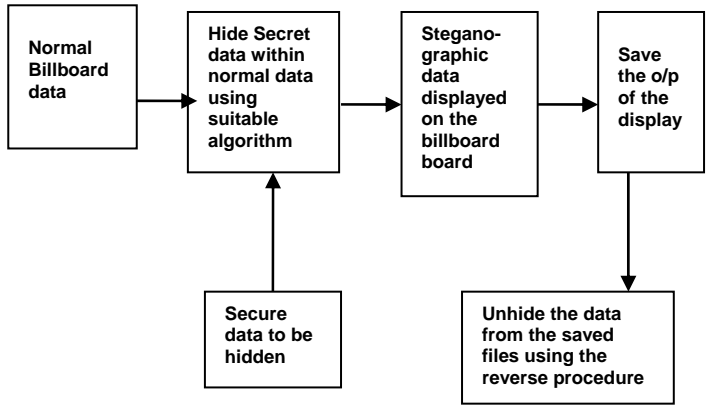

Figure 2 Block diagram of the proposed idea

a) Send the normal data that has to be displayed to the display board.
b) Using a suitable Steganography algorithm hide the secret data within the normal data before sending it to the display board.

This method can be used for announcing a secret message in public place. It can be extended to other means such as electronic advertising board around sports stadium

A. *Description of the algorithm for embedding the secret message:*

Algorithm for embedding the secret message is as follows:

a) Read the image from the source.
b) Divide the image into [R x C] smaller blocks .Where R & C are the first & second bytes of the key respectively [Figure 4].
c) Each smaller block is a combination of many pixels of different values.
d) The LSBs of the pixel are changed depending on the pattern bits and the secret message bits.
e) The pattern bits are considered in sequence form its MSB.
f) If the pattern bit is 0, then the first LSB of the pixel is changed [i.e if data bit is 1 and pixel bit is 0 ,then pixel bit is changed to1 or else it is retained as it is.]

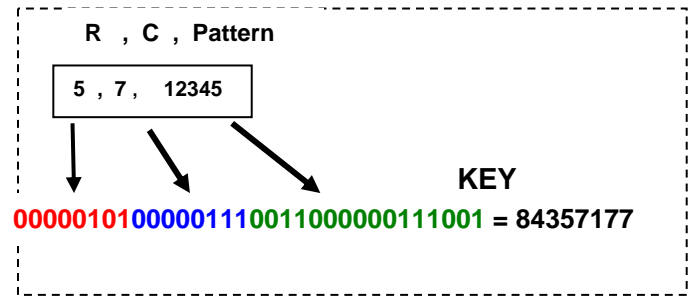

g) If the pattern bit is 1, then the second LSB of the pixel is changed accordingly.
h) A single bit of the secret message is distributed through out the block. This is done to have enough information so that correct information can be retrived after decoding
i) Similarly the other bits are inserted in the remaining blocks.
j) If the length of the secret message is large , then it can be divided and stored in two or three frames.
k) To extract the information, operations contrary to the ones carried out in embedding are performed.

The key plays a very important in embedding the message. Larger the key size, the more difficult to suspect the secrecy. The figure 3 shows how the key is generated. The first 8 bits in red colour represent the no. of rows R & the next 8 bits in blue colour represents the no. of columns C. The 16 bits in green followed by row and column represent the pattern bits. Each whole block of the cover image includes only one bit of the secret data. This is done so that more amount of data is available during retrieval.

B. *Performance Measures*

The performance measure depends on the success rate of the implementation of the overall system with respect to the following points.

a) The integrity of the hidden information should not change after embedding.

b) The stego object must remain almost remain unchanged to the naked eye.
c) There should be accuracy in the extracted data

V RESULTS

In order to demonstrate the Online transmission of the hidden data, 3 systems are used
System 1 : To create and send the normal billboard data ( any advertisement )
System 2 : To hide the secret message .
System 3 : To display any data coming from system 2.

Four modules are designed for the same in as shown in figures 4,5,6,7 below.

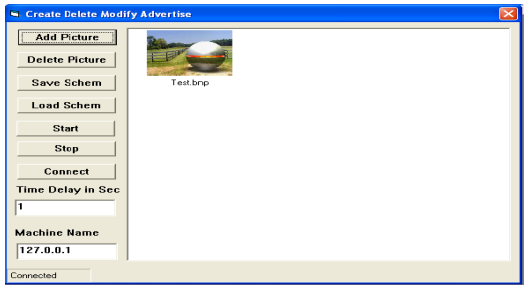

Figure 4. Module to create, delete & modify the advertisement

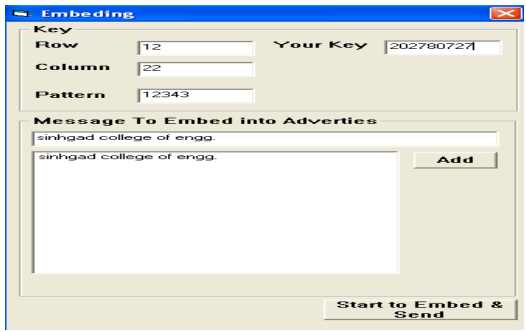

Figure 5. Module to embed the secret message

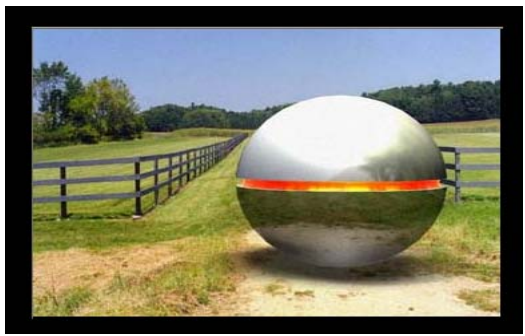

Figure 6. LCD display

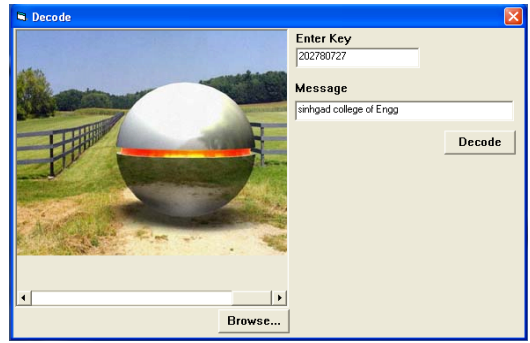

Figure 7 . Module to decode the secret information

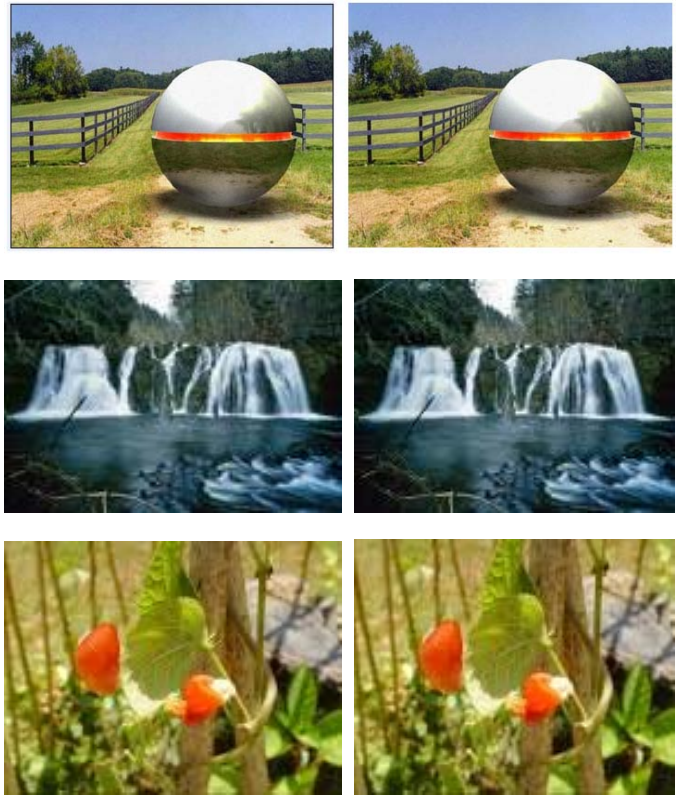

Figure 8a. Original images    8 b.    Stego images

The Module (create ,delete & modify advertisemenr) is basically used to create a data base for the advertisement and send it to the second system ,where the data is to be hidden. The stego image from the second system is sent to the display board . This image is captured by the camera & the decode module decodes the given information. The results shown consist of the image that is directly saved on to the desktop .

## VI    Conclusion

As steganography becomes more widely used in computing, there are issues that need to be resolved. There are a wide variety of different techniques with their own advantages and disadvantages. Many currently used techniques are not robust enough to prevent detection and removal of embedded data. The use of benchmarking to evaluate techniques should become more common and a more standard definition of robustness is required to help overcome this. For a system to be considered robust it should have the following properties:

a) The quality of the media should not noticeably degrade upon addition of a secret data.
b) Secret data should be undetectable without secret knowledge, typically the key.
c) If multiple data are present they should not interfere with each other.
d) The secret data should survive attacks that don't degrade the perceived quality of the work.

This work presents a scheme that can transmit large quantities of secret information and provide secure communication between two communication parties. Both steganography and cryptography can be woven into this scheme to make the detection more complicated. Any kind of text data can be employed as secret msg. The secret message employing the concept of steganography is sent over the network .In addition, the proposed procedure is simple and easy to implement. Also, the developed system has many practical, personal and militaristic applications for both point-to-point and point-to-multi-point communications